\UseRawInputEncoding

\documentclass[aps,showpacs,twocolumn,pra,superscriptaddress]{revtex4}
\usepackage[tbtags,fleqn]{amsmath}
\usepackage{amssymb}
\usepackage{amsfonts}
\usepackage{mathtools}
\usepackage[colorlinks=true,citecolor=blue,urlcolor=blue]{hyperref}
\hypersetup{colorlinks,linkcolor=blue,filecolor=blue,urlcolor=blue,citecolor=blue}
\usepackage{multirow}
\usepackage{graphicx}
\usepackage{subfigure}
\usepackage{cases}

\begin{document}
	\title{ Entanglement-Enhanced Quantum Strategies for Accurate Estimation of Multibody Group Motion and Moving Object Characteristics}
	
	\author{Yongqiang Li}
	\affiliation{Key Laboratory of Low-Dimensional Quantum Structures and Quantum Control of Ministry of Education, Key Laboratory for Matter Microstructure and Function of Hunan Province, Department of Physics and Synergetic Innovation Center for Quantum Effects and Applications, Hunan Normal University, Changsha 410081, China}
	\author{Changliang Ren}\thanks{Corresponding author: renchangliang@hunnu.edu.cn}
	\affiliation{Key Laboratory of Low-Dimensional Quantum Structures and Quantum Control of Ministry of Education, Key Laboratory for Matter Microstructure and Function of Hunan Province, Department of Physics and Synergetic Innovation Center for Quantum Effects and Applications, Hunan Normal University, Changsha 410081, China}
	
	\begin{abstract}
		This study presents a quantum strategy for simultaneous estimation of two physical quantities using different entanglement resources. We explore the utilization of positively or negatively time-correlated photons. The proposed method enables the detection of central position and relative velocity of multibody systems, as well as precise measurement of size and velocity of moving objects. Comparative analysis with other strategies reveals the superior quantum advantage of our approach, particularly when appropriate entanglement sources with high entanglement degree are employed. These findings contribute to advancing our understanding of quantum strategies for accurate measurements.\par
	\end{abstract}

	
	\maketitle
	
	\section{Introduction}  $\label{introduction}$
	Quantum metrology \cite{PhysRevLett.96.010401,giovannetti2011advances,szczykulska2016multi,QuantumMetrologyPhysRevLett.107.083601,OpticalQuantumMetrologyPRXQuantum.3.010202,huang2016usefulness} is an emerging application of quantum information technology that follows quantum communication \cite{gisin2007quantum} and quantum computing \cite{AndrewSteane1998,quantumcomputing}, which aims to achieve higher precision measurements using quantum strategies. Along with the latest technological advances in quantum optics, electricity, and opto-mechanical vibronic systems, quantum metrology has been applied to many practical tasks, which have significantly boosted the development of related fields, such as gravitational wave detection \cite{PhysRevLett.116.061102,Quantummetrologygravitationalwave}, force sensing \cite{degen2017quantum,Zhang_2021}, magnetic force measurement \cite{geremia2003quantum}, clocks \cite{giovannetti2001quantum,giovannetti2004conveyor}, and biological measurements \cite{DARIANO2005133,Quantummetrologybiology}.\par
	
	As the fundamental physical quantities, the precision measurement of time and frequency is an essential ingredient of quantum metrology \cite{PhysRevLett.96.010401,giovannetti2011advances}. According to these studies, enhancing the detection accuracy of radar becomes an important potential application. Several quantum radar schemes have been proposed, such as quantum illumination \cite{PhysRevLett.101.253601,PhysRevLett.114.080503,shapiro2020quantum,nair2020fundamental,guha2009gaussian,zhuang2022ultimate}, quantum positioning \cite{giovannetti2001quantum}, and 3D accuracy-enhanced quantum radar  \cite{PhysRevLett.124.200503,9477766} etc, some of which have also had demonstrative experiments \cite{lopaeva2013experimental,PhysRevLett.127.040504}. Most of the current research can be attribute to a single parameter estimation problem, such as determining positions or velocities. The ultimate measurement accuracy of them that can be achieved is determined by the Quantum Cramér-Rao bound (QCRB) \cite{szczykulska2016multi,demkowicz2020multi,liu2019quantum,paris2009quantum}, which is obtained by analyzing the Quantum Fisher Information (QFI) \cite{liu2019quantum,szczykulska2016multi,demkowicz2020multi,paris2009quantum}. Compared to classical strategies, quantum strategies can improve accuracy and sensitivity, where the maximum accuracy achievable for a single parameter is the Heisenberg limit which transcends the standard quantum limit. However, in various application scenarios, the positioning of multibody system, and the size of the object, we actually need to determine different physical quantities simultaneously, such as the simultaneous estimation of central position and relative velocity. Although quantum multiparameter estimation has started to be explored \cite{demkowicz2020multi,liu2019quantum}, few generalized quantum positioning schemes were designed based on quantum multiparameter estimation theory. With respect to multiparameter estimation, subject to the incommute relation between different observables, it often exists a tradeoff, i.e., Heisenberg's uncertainty principle \cite{Heisenberg,BUSCH_2007}. Zhuang \emph{et.al} proposed entanglement-induced lidar for measuring a target’s range and velocity \cite{zhuangPhysRevA.96.040304}. Recently, Huang \emph{et.al }shows that the tradeoff in simultaneous estimating both time and frequency can be weakened when uses entangled states as probe states \cite{PRXQuantum.2.030303}. In this study, we introduce a quantum strategy aimed at simultaneously estimating two distinct physical quantities in various application scenarios utilizing different quantum entanglement resources. Specifically, we investigate the utilization of photons with positive time correlation or negative time correlation.
	
	Firstly, we propose a method for detecting the central position and relative velocity of a multibody system by measuring the time sum and frequency difference. Secondly, we present a technique for precisely measuring the size and velocity of moving objects. To provide a comprehensive analysis, we also delve into the accuracy limits of the aforementioned estimation parameters when employing the quantum illumination strategy and the single-photon strategy mentioned earlier. Our results demonstrate that when appropriate entanglement sources are utilized and the entanglement degree is relatively high, this strategy exhibits a superior quantum advantage compared to the other two strategies.
	
	The article is structured as follows: In the section \ref{II}, we delve into the estimation of the central position and relative velocity of two-object systems. We employ quantum multi-parameter estimation theory to establish a comprehensive method for analyzing the measurement accuracy limits across different schemes. The section \ref{III} introduces quantum strategies specifically designed for measuring the size and velocity of objects. We outline approaches employed in this context. Finally, Section \ref{IV} presents the concluding remarks and summarizes the key findings and implications discussed throughout the article.

	\begin{figure}[htbp]
		\centering
		\includegraphics[width=0.5\textwidth,height=0.3\textwidth]{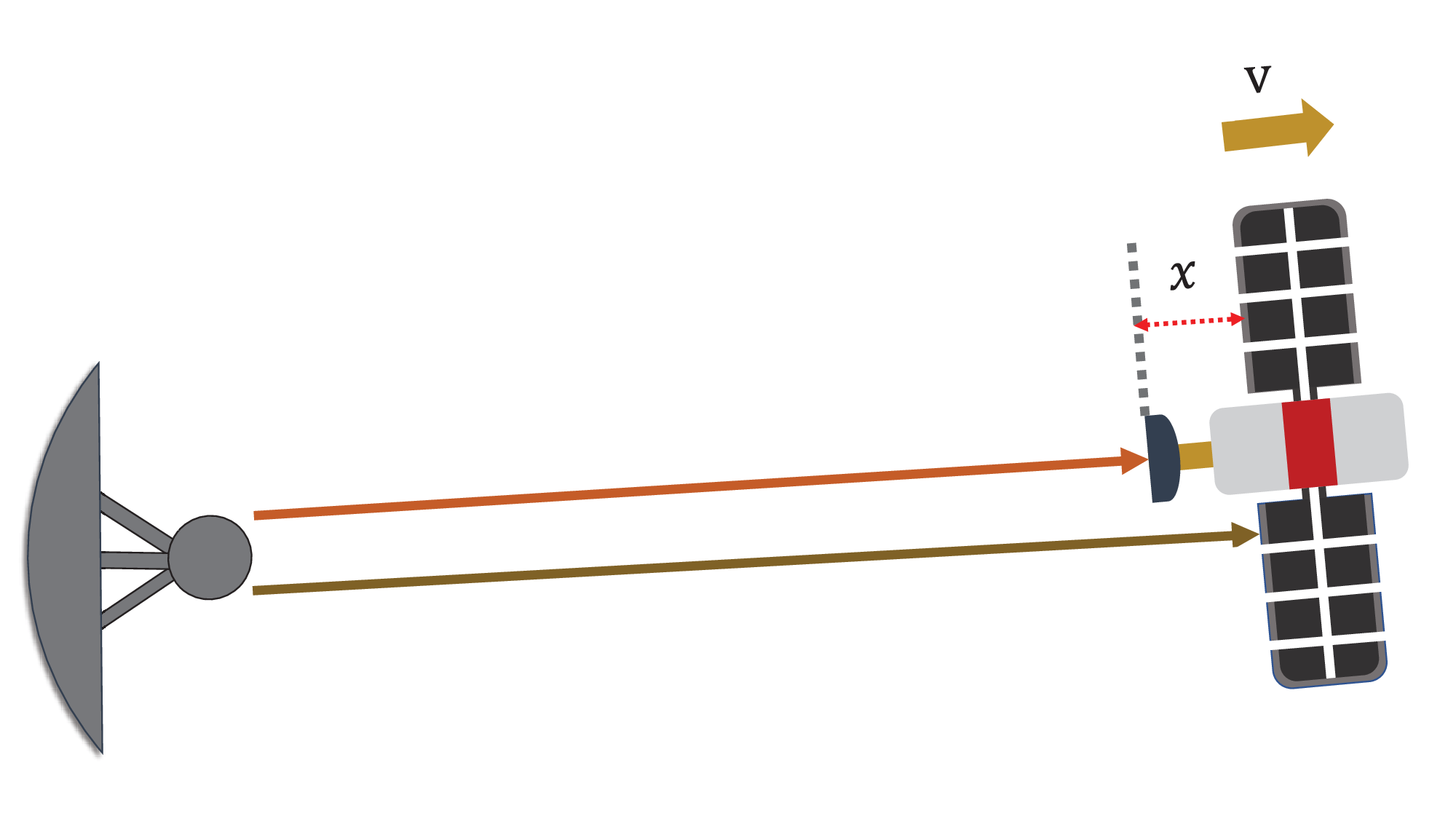}
		\caption{Illustration of Central Position and Relative Velocity Detection in a Multibody System.}\label{figure}
	\end{figure}	
	
	\section{Estimation of group motion of a multibody system} \label{II}

	We present a method for detecting the central position and relative velocity of a multibody system through the measurement of the time sum and frequency difference. To describe the concept, let us examine a basic multibody system consisting of two objects as an illustrative example. As illustrated in Fig.\ref{figure}, supposed that
	two photons arrive at two objects moving with velocities $v_{1}$ and $v_{2}$ after travel times $\frac{\bar{t}_{1}}{2}$ and $\frac{\bar{t}_{2}}{2}$. The position $\vec{r}_{1}=\vec{c}_{1} \frac{ \bar{t}_{1}}{2}$ of object A, as well as the position $\vec{r}_{2}=\vec{c}_{2} \frac{ \bar{t}_{2}}{2}$ of object B, can be measured. Hence the central position between these two object can be expressed by
	\begin{eqnarray}
		\vec{r}=\frac{\vec{c}_{1}\bar{t}_{1}\label{distance} +\vec{c}_{2}\bar{t}_{2}}{2},
	\end{eqnarray}
	where the $\vec{c}_{1}$ and $\vec{c}_{2}$ are two velocity vectors of the returned photons. The two objects are located at a considerable distance from the radar, $\vec{c}_{1}\approx\vec{c}_{2}$, enabling their central position to be precisely determined as $\vec{r}=\frac{\vec{c}(\bar{t}_{1} +\bar{t}_{2})}{2}$. Obviously, the accuracy of determining the central position relies on the variance of $\bar{t}_{1} + \bar{t}_{2}$. Besides, the relative velocity of two objects can also be estimated by the Doppler effect. The relationship between the difference in frequency and the difference in velocity can be described as follows,
	\begin{eqnarray}\label{frequency shift}
		&\bar{\omega }_{2}-\bar{\omega }_{1}=  \nonumber (\frac{c-v_{2} }{c+v_{2} }\bar{\omega } _{0} -\bar{\omega }_{0})-(\frac{c-v_{1} }{c+v_{1} }\bar{\omega } _{0} -\bar{\omega }_{0}) \\&\simeq  \frac{2(v_{2}-v_{1})  }{c} \bar{\omega }_{0}
	\end{eqnarray}
	where $\bar{\omega } _{0}$ is the central frequency of the initial pulse, $\bar{\omega }_{1}$ and $\bar{\omega }_{2}$ are the central frequencies of the returned photons from different direction. Similarly, the accuracy of determining  the relative velocity of two objects relies on the variance of $\bar{\omega }_{2}-\bar{\omega }_{1}$.

	To accomplish the task, we initially utilize a 2-qubit entangled state where two photons are emit to these two objects respectively. In this context, we are specifically focusing on the ideal scenario where the direction is precisely defined, and the photons experience highly efficient reflection. Without loss of generality, the initial emitted state can be described in the time domain,
	\begin{eqnarray}\label{entangled}
		\left | \psi  \right \rangle =\int \int \phi _{0} (t_{1},t_{2}  )\left | t_{1}   \right \rangle \left | t_{2}   \right \rangle dt_{1}dt_{2}, \label{3}
	\end{eqnarray}
	where $\left |t _{i}   \right \rangle=\hat{a}^{\dagger }  (t _{i} )\left | 0  \right \rangle $ is a single-photon state at $t_{i}$ with $\{i=1,2\}$, and $\hat{a}^{\dagger}  (t _{i} )$ is the creation operator. $\phi _{0} (t_{1},t_{2}  )$ is biphoton temporal wavefunction. We choose $\phi _{0}(t_{1},t_{2}  )=\sqrt{\frac{2\sqrt{1-\kappa^2}\sigma_{0} ^2 }{\pi } }e^{-\sigma_{0} ^2(t_{1}   ^2+t_{2}   ^2-2\kappa t_{1}t_{2})}  e^{-i\bar{\omega} _{0}(t_{1}+t_{2})  }$, where $\bar{\omega} _{0}$ and $\sigma_{0}$ are the carrier frequency and bandwidth for each single photon. The
	parameter $\kappa\in (-1,1)$ quantifies the amount of entanglement between the two photons. As $\kappa$ approaches 1, the biphoton state is positively correlated in time. While the two photons will negatively correlated in time when $\kappa$ approaches $-1$. And $\kappa=0$ means they are separate state.
	
	The two photons of the biphoton state experience backscattering and subsequently exhibit a time delay upon their return, where a detailed derivation can be found in \cite{PRXQuantum.2.030303} (also refer to Appendix \ref{appendix B}).
	Without loss of generality, the returned biphoton state can be expressed as,
	\begin{eqnarray}
		\left | \psi^{'}  \right \rangle =\int \int \phi (t_{1},t_{2}  )\left | t_{1}   \right \rangle \left | t_{2}   \right \rangle dt_{1}dt_{2},  \label{4}
	\end{eqnarray}
	where $\phi(t_1,t_2) = \sqrt{\frac{2\sqrt{1-\kappa^2}\sigma_{1}\sigma_{2}}{\pi } }e^{-i\bar{\omega }_{1}(t_{1}-\bar{t}_{1}   )  }e^{-i\bar{\omega} _{2}(t_{2}-\bar{t}_{2}   ) } \\  e^{-[(t_{1} -\bar{t}_{1}  )^2\sigma_{1} ^{2}+(t_{2}-\bar{t}_{2})^2\sigma_{2} ^{2}-2\kappa \sigma_{1}\sigma_{2} (t_{1}-\bar{t}_{1})(t_{2}-\bar{t}_{2}) ]}$ with
	\begin{flalign}
		\begin{aligned}
			& \ \sigma_{i} = \frac{c-v_{i} }{c+v_{i} }\sigma_{0}
			& \ &
			\bar{t}_{i}  =  \frac{2r_{i}}{(c-v_{i} )}
			&& \\
			& \
			\bar{\omega}_{1}    =  \frac{c-v_{1} }{c+v_{1} } \bar{\omega}_{0}
			& \ & \bar{\omega}_{2}    =  \frac{c-v_{2} }{c+v_{2}} \bar{\omega}_{0}
			&&
		\end{aligned}.\label{5}
	\end{flalign}

	As the goal is to compute the QFI matrix for the estimation of the central position and relative velocity, according to the Eqs.(\ref{distance}) and (\ref{frequency shift}), it is better to define these
	variables $\bar{\omega}_{-}    =   \bar{\omega}_{2}-\bar{\omega}_{1},\bar{t}_{+}  =  \bar{t}_{1} +\bar{t}_{2},\bar{\omega}_{+}    =   \bar{\omega}_{1}+\bar{\omega}_{2},\bar{t}_{-}  =  \bar{t}_{2} -\bar{t}_{1}$. After performing detailed and coherent calculations (refer to Appendix \ref{appendix B} and Appendix \ref{appendix C}), we derive a precise expression for the QFI matrix, allowing us to accurately estimate the estimation $\bar{t}_{+}$ and $\bar\omega_{-}$,
	\begin{eqnarray}\label{QFI for entangle}
		H(\bar{t}_{+},\bar\omega_{-}) & = & \begin{pmatrix}
			2(1-\kappa )\sigma ^2 &0 \\
			0&  \frac{1}{2(1+\kappa )\sigma^2} \\
		\end{pmatrix}.
	\end{eqnarray}
	
	According to the necessary and sufficient condition for joint optimal estimation \cite{paris2009quantum}, we find $\mathrm{Tr}[\rho[L_{\bar{t}_{+}},L_{\bar{\omega}_{-}}]]=0$ where $L_{\lambda}$ the symmetric logarithmic
	derivative matrix (SLD). Therefore, by utilizing an entangled photon state that exhibits negatively temporal correlation as the emission source, we can achieve the saturation of the QCR bound \cite{szczykulska2016multi,demkowicz2020multi,liu2019quantum,paris2009quantum}, indicating the estimation of the time sum and the frequency difference can be achieved optimally.
	Obviously, the variance of the estimator $\bar{t}_{+}$ depends on the reciprocal of $2(1-\kappa )\sigma ^2$, while the variance of the estimator $\bar\omega_{-}$ depends on the reciprocal of $\frac{1}{2(1+\kappa )\sigma^2}$. These two tems can be arbitrarily small simultaneously. The relation of estimating $\bar{t}_{+}$ and $\bar\omega_{-}$ can be expressed as
	\begin{eqnarray}
		\delta \bar{t}_{+}\delta \bar\omega_{-}\ge \frac{\sqrt{1+\kappa}}{\sqrt{1-\kappa}}.\label{7}
	\end{eqnarray}
	Clearly, when $\kappa\rightarrow -1$, the right side of this relation tends towards zero, indicating that the tradeoff in precision between $\bar{t}_{+}$ and $\bar\omega_{-}$ can be alleviated. As $\kappa$ deviates from $-1$, the accuracy of joint estimation deteriorates progressively.
	In fact, the accuracy of joint estimation of $\bar{t}_{+}$ and $\bar\omega_{-}$ directly influences the accuracy of estimating the central position and the relative velocity. This analysis is conducted directly and discussed in detail in the Appendix \ref{appendix A}.
	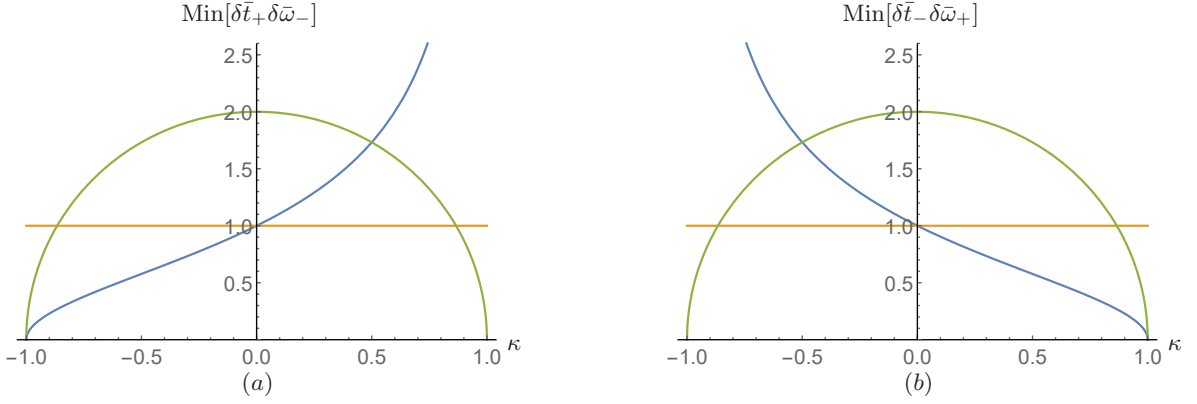
\begin{figure*}[!htbp]
		\begin{picture}(480,160)
			
			\put(50,10){\makebox(120,135){
					\scalebox{0.4}[0.4]{\includegraphics{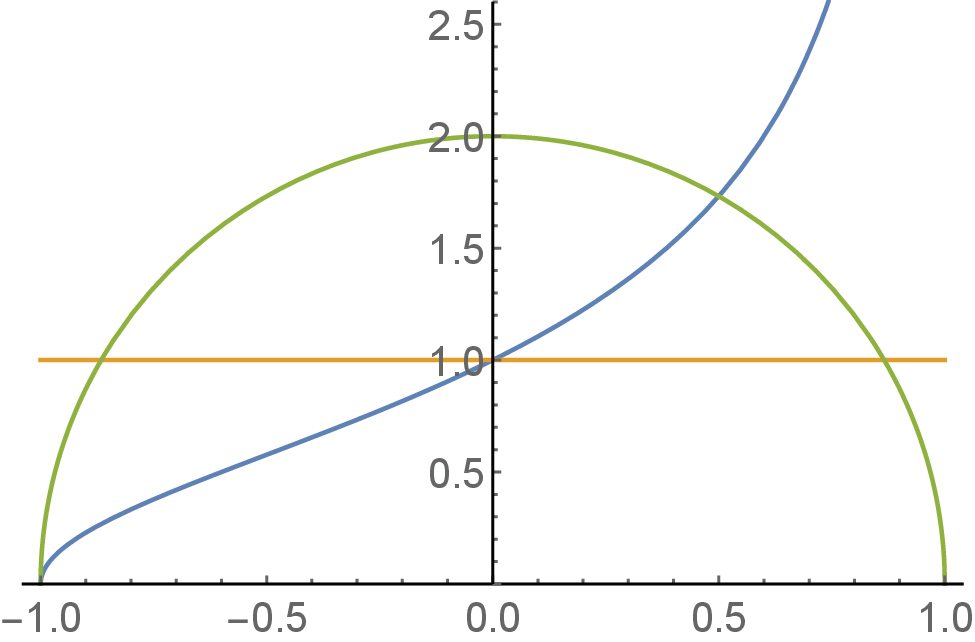}}}}

			\put(300,10){\makebox(120,135){
					\scalebox{0.4}[0.4]{\includegraphics{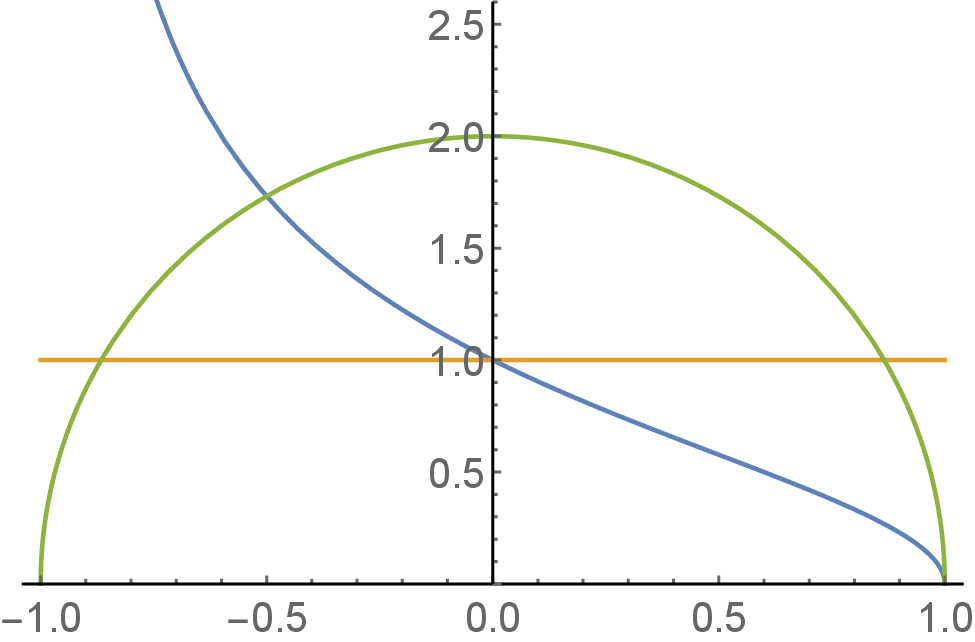}}}}

			\put(100,100){\makebox(20,100){Min[$\delta\bar{t}_{+}\delta\bar{\omega}_{-}$]}}
			\put(350,100){\makebox(20,100){Min[$\delta\bar{t}_{-}\delta\bar{\omega}_{+}$]}}
			\put(450,-25){\makebox(20,100){$\kappa$}}
			\put(200,-25){\makebox(20,100){$\kappa$}}
			\put(103,-40){\makebox(20,100){$(a)$}}
			\put(353,-40){\makebox(20,100){$(b)$}}
		\end{picture}
		\caption{The figure (a) shows the accuracy limit of the simultaneous estimation of time sum and frequency difference for different strategies. The figure (b) shows the accuracy limit of the simultaneous estimation of time difference and frequency sum for different strategies. The blue line, green line and yellow line correspond to our strategy, quantum illumination strategy and single photon strategy respectively.}\label{figure 2}
	\end{figure*}

	Compared with biphoton state, we can initially utilize two single-photon states as emitted resource where two photons emit to these two
	objects independently. Each single-photon state can be described as $\left | \psi_i \right \rangle =\int \psi_{0}({t_i}) \left | t_i \right \rangle \mathrm{d}t_i$ initially, where the temporal distribution is
	$\psi_0(t_i)  = ({\frac{2\sigma _{0}^2}{\pi }})^{\frac{1}{4} }e^{-t_i^2\sigma_{0} ^{2}}e^{-i\bar{\omega }t_i}$ with same spectrum $\phi_0(t_1,t_2)$ as in (\ref{entangled}). These two photons experience
	backscattering and subsequently exhibit a time delay upon their return, the density matrix of the returned two single-photon states can be described by \cite{PhysRevX.6.031033,PRXQuantum.2.030303}
	\begin{eqnarray}
		\rho_{1}  =  (\lvert  \psi_{1}  \rangle  \langle \psi_{1} \rvert  +\lvert  \psi_{2} \rangle  \langle \psi_{2} \rvert) \label{8}
	\end{eqnarray}
	where $\lvert  \psi_{i} \rangle  = \int \psi _{i}(t_i)  \lvert  t_{i}  \rangle dt_{i}$, and the temporal distribution is $\psi _{i} (t_i) = ({\frac{2\sigma _{i}^2}{\pi }})^{\frac{1}{4} }e^{-(t_{i} -\bar{t}_{i}  )^2\sigma_{i} ^{2}}e^{-i\bar{\omega }_{i}(t_{i}-\bar{t}_{i})}$. It is important to note that, in order to compare with other strategies using the same photon number, Eq.(\ref{8}) is not normalized.
	
	As illustrated in Appendix \ref{appendix C}, the QFI matrix that allows us to accurately estimate the estimators $\bar{t}_{+}$ and $\bar\omega_{-}$, is given by,
	\begin{eqnarray}
		H  =  \begin{pmatrix}
			H_{\bar{t}_{+}^2} & 0\\
			0 &  H_{\bar{\omega}_{-}^2}
		\end{pmatrix},
	\end{eqnarray}
	where $H_{\bar{t}_{+}^2}=2\sigma^2+\varepsilon _{\bar{t} _{+} }>0 $,
	$H_{\bar{\omega}_{-}^2}=\frac{1}{2\sigma^2}+\varepsilon _{\bar{\omega} _{-} }>0$,
	$\varepsilon _{\bar{t}_{+} }\le0$, $\varepsilon _{\bar{\omega}_{-}}\le0$ (Appendix \ref{appendix C} for detailed expressions of $\varepsilon _{\bar{t}_{+}}$ and $\varepsilon _{\bar{\omega}_{-}}$). Clearly, we find $H_{\bar{t}_{+}^2}\le 2\sigma^2$ and $H_{\bar{\omega}_{-}^2}\le\frac{1}{2\sigma^2}$. Likewise, the estimation of the time sum and the frequency difference can be achieved optimally because $\mathrm{Tr}[\rho[L_{\bar{t}_{+}},L_{\bar{\omega}_{-}}]]=0$.
	Hence, the joint estimation for $\bar{t}_{+}$ and $\bar\omega_{-}$ satisfies
	\begin{eqnarray}
		\delta \bar{t}_{+}\delta \bar{\omega}_{-}\ge  1.\label{10}
	\end{eqnarray}
	Obviously, due to entanglement, the biphoton state can achieve much lower bound in terms of joint measurement accuracy of time sum and frequency difference, and this advantage will be lost when $\kappa \ge 0$. \par
	
	Likewise, we can also measure the multibody relative velocity as well as central position using the quantum illumination-based scheme, as discussed by Huang et al \cite{PRXQuantum.2.030303}. To accomplish the task, we need two pairs of entangled photons, the signal photon of each pair will emit to the two objects respectively, while their idler photons remain. By jointly measuring each scattered signal photon and the corresponding idler photon, the distance and velocity of each object can be obtained from the time delay and frequency shift of the scattered signal photon respectively. Hence the relative velocity and central position of this multibody system are estimated by firstly measuring the distance and velocity information of each object. Therefore, this scheme is actually similar to the single-photon scheme,
	the density matrix of the used state can be expressed as $\rho_{2} =\frac{1}{2} (\lvert \Psi_{1} \rangle \langle\Psi_{1} \rvert +\lvert \Psi_{2}\rangle \langle\Psi_{2} \rvert) $, where $\lvert \Psi_{i} \rangle$ is a biphoton entangled state as defined Eq.(\ref{4}). Unlike the single-photon scheme, the tradeoff in simultaneous estimating both time and frequency can be weakened when uses entangled states as probe states. Similarly, we can derive the QFI matrix that allows us to accurately estimate the estimators $\bar{t}_{+}$ and $\bar\omega_{-}$, which is $H  = \begin{pmatrix}
		H_{\bar{t}_{+}^2} & 0\\
		0 &  H_{\bar{\omega}_{-}^2}
	\end{pmatrix}$
	where $H_{\bar{t}_{+}^2}=\sigma^2+\varepsilon _{\bar{t} _{+} } $, $H_{\bar{\omega}_{-}^2}=\frac{1}{4(1-\kappa^2)\sigma^2}+\varepsilon _{\bar{\omega} _{-} }$. In this scheme, the joint estimation for $\bar{t}_{+}$ and $\bar\omega_{-}$ satisfies
	\begin{eqnarray}
		\delta \bar{t}_{+}\delta \bar{\omega}_{-}\ge  2\sqrt{1-\kappa^2}.
	\end{eqnarray}
	It is imply that the uncertainty of simultaneously estimating $\bar{t}_{+}$ and $\bar\omega_{-}$ can be reduced by factor $2\sqrt{1-\kappa^2}$ using this scheme.\par
	
	As depicted in Fig.\ref{figure 2}(a), we can compare the estimation accuracy of the central position and relative velocity for various types of detection sources and strategies. To ensure fairness, we maintain consistency by employing an equal number of photons for different probing strategies. The lower bound of the uncertainty of simultaneously estimating $\bar{t}_{+}$ and $\bar\omega_{-}$, denoted as $\mathrm{Min}[\delta \bar{t}_{+}\delta \bar{\omega}_{-}]$, are the function of the parameter $\kappa$ for different strategies. In our strategy, we achieve $\mathrm{Min}[\delta \bar{t}_{+}\delta \bar{\omega}_{-}]=\frac{\sqrt{1+\kappa}}{\sqrt{1-\kappa}}$ (blue), while the quantum illumination scheme yields $\mathrm{Min}[\delta \bar{t}_{+}\delta \bar{\omega}_{-}]=2\sqrt{1-\kappa^2}$ (green). In the single-photon strategy, $\mathrm{Min}[\delta \bar{t}_{+}\delta \bar{\omega}_{-}]$ remains constant at $1$ throughout.

	It is clearly shown that, when utilizing a time-negatively correlated entanglement source for estimation, $\kappa\in(-1,0)$, our scheme proves to be optimal. However, for a time-positively-correlated entanglement source, the situation varies depending on the quality of the entanglement. Our strategy and the quantum illumination strategy fall short of the performance achieved by the single-photon strategy, when the entanglement is relatively weak, $\kappa\in(0,\frac{\sqrt{3}}{2})$. Conversely, when the entanglement source approaches ideal positively correlated in time, $\kappa\in(\frac{\sqrt{3}}{2},1)$, the the quantum illumination strategy emerges as the optimal solution.
	
	\begin{figure}[htbp]
		\centering
		\includegraphics[width=0.48\textwidth,height=0.27\textwidth]{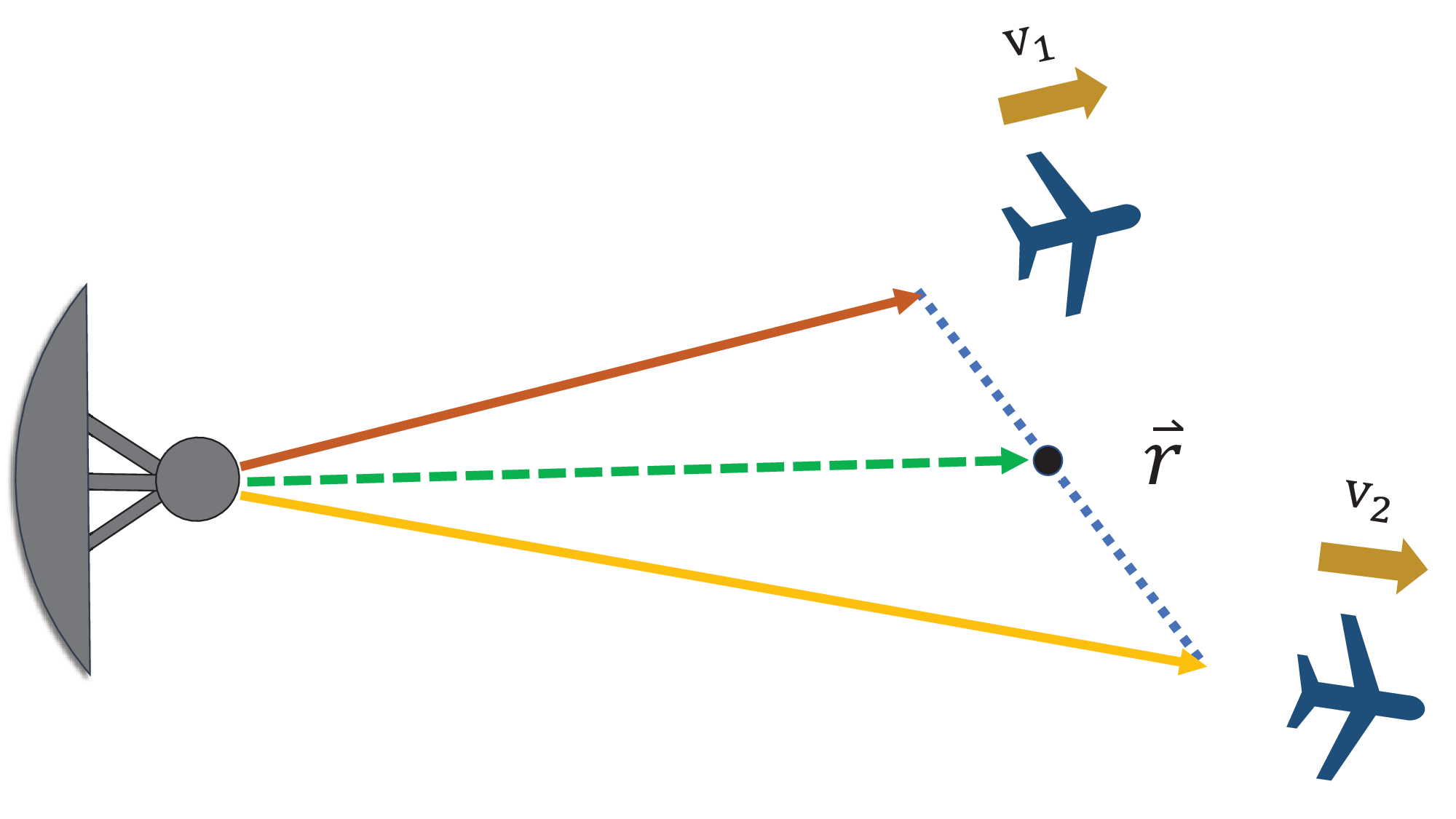}
		\caption{Illustration of measuring the relative size and velocity of moving object. } \label{figure 3}
	\end{figure}
	
	\section{Measuring the relative size and velocity of moving object} \label{III}
	
	We present a method to precisely measure the size and velocity of moving objects. In our scenario, we focus on measuring a far-field moving target with a velocity along the detection direction. The targets have well-defined sizes and radial distributions. We are particularly interest in measuring the relative radial distance between two distinct regions of the target and the velocity of the target. By iteratively performing this procedure, we can achieve comprehensive imaging of the target. As illustrated in Fig.\ref{figure 3}, the object moves along the radial direction of the radar with velocity $v$.
	
	Assuming that the distance between two distinct points on an object, labeled as A and B, and the radar are $R_{0}$ and $R_{0}+x$, respectively, where $x$ represents the relative difference in size between the two points. To accomplish this task, a minimum of two photons is required. When the radar emits two photons simultaneously that are backscattered by the two distinct points of the object repectively, each photon will return with a travel time $\bar{t}_{i}$ and a central frequency $\bar{\omega}_{i}$. According to Eq.(\ref{5}), it is easy to obtain that $\bar{t}_{1} = \frac{2R_{0}}{c-v}$ and $\bar{t}_{2} = \frac{2R_{0}+2x}{c-v}$. Hence the time difference caused by the size of the object is $\bar{t}_{2}-\bar{t}_{1} = \frac{2x}{c-v}\approx\frac{2x}{c}$, and the distance between the two distinct point can be given as
	\begin{eqnarray}
		x=\frac{c(\bar{t}_{2}-\bar{t}_{1})}{2}.
	\end{eqnarray}
	Clearly, the accuracy of determining the distance $x$ relies on the variance of $\bar{t}_{2} - \bar{t}_{1}$. Likewise, the velocity of the object can be estimated by analyzing the frequency shift of the returned photons. According to the Doppler effect, the carrier frequency of the returned photon will change as $\bar{\omega}_{i} = \frac{c-v}{c+v} \bar{\omega}_{0}$, so the frequency shift determines $\bigtriangleup \omega_{i} = \bar{\omega}_{i} - \bar{\omega }_{0} \approx\frac{2\bar{\omega}_{0}v}{c}$. Although it is possible to measure velocity using a single photon, at least two photons is necessary for the complete task especially for measuring the size of the object. Hence, we analyze the process of employing two photons to measure the velocity of an object. At first glance, this may seem like a trivial process. Assuming that two photon pulses have identical initial central frequencies and the velocity of each part of the object is the same, the frequency shift of the returned
	photons will satisfy $\bigtriangleup \omega_{1}=\bigtriangleup \omega_{2}=\bigtriangleup \omega$, The velocity of the object can be estimated by summing up the frequency shifts of the two photons, which can be given as,
	\begin{eqnarray}
		v=\frac{c}{2\bar{\omega}_{0}}(\bar{\omega}_{1}+\bar{\omega}_{2}-2\bar{\omega}_{0})=\frac{c\bigtriangleup \omega}{\bar{\omega}_{0}}.
	\end{eqnarray}
	Hence, the accuracy of determining the velocity relies on
	the variance of $\bar{\omega}_{1}+\bar{\omega}_{2}$. It is possible to reduce this variance to infinitesimal levels using specialized photon states, thereby enhancing the precision of measuring velocity.\par
	Likewise, employing a two-photon entangled state Eq.(\ref{3}) as the emission source directed towards the object, and subsequently the photons reflected back, the returned biphoton state will change as Eq.(\ref{4}). After performing detailed and coherent calculations (refer to Appendix \ref{appendix B} and Appendix \ref{appendix C}), we derive a precise expression for the QFI matrix, allowing
	us to accurately estimator $\bar{t}_{-}$ and $\bar{\omega}_{+}$,
	\begin{eqnarray}
		H(\bar{t}_{-},\bar{\omega}_{+}) & = &
		\begin{pmatrix}
			2(1+\kappa )\sigma ^2 &0 \\
			0&  \frac{1}{2(1-\kappa )\sigma ^2} \\
		\end{pmatrix}.
	\end{eqnarray}
	We compute the necessary and sufficient condition, $\mathrm{Tr}[\rho[L_{\bar{t}_{-}},L_{\bar{\omega}_{+}}]]=0$, indicating that the parameters $\bar{t}_{-}$ and $\bar{\omega}_{+}$ can be joint optimal estimation. Obviously, the variances of the $\bar{t}_{+}$ depends on the reciprocal of $2(1+\kappa )\sigma ^2$, while the variances of the $\bar\omega_{-}$ depends on the reciprocal of $\frac{1}{2(1-\kappa )\sigma^2}$. Therefore, we can obtain
	\begin{eqnarray}
		\delta \bar{t}_{-}\delta \bar{\omega}_{+}\ge \frac{\sqrt{1-\kappa}}{\sqrt{1+\kappa}}.
	\end{eqnarray}\par
	Notably, when $\kappa\rightarrow1$, the right side of this relation tends towards zero. As $\kappa$ deviates from 1, the accuracy of joint estimation deteriorates progressively.\par
	For the probing state being two single-photon states, according to the photon's travel time and frequency shift, the photon state returned can be expressed as Eq.(\ref{8}).
	As illustrated in Appendix \ref{appendix C}, the QFI matrix that allows us to accurately estimate the estimation $\bar{t}_{-}$ and $\bar\omega_{+}$, is
	\begin{eqnarray}
		H  =  \begin{pmatrix}
			H_{\bar{t}_{-}^2} & 0\\
			0 &  H_{\bar{\omega}_{+}^2}
		\end{pmatrix},
	\end{eqnarray}
	where $H_{\bar{t}_{-}^2}=2\sigma^{2}+\varepsilon _{\bar{t} _{-} }>0$ and $H_{\bar{\omega}_{+}^2}=\frac{1}{2\sigma^2}+\varepsilon _{\bar{\omega} _{+} }>0$, with $\varepsilon _{\bar{t}_{-}}\le0$ and $\varepsilon _{\bar{\omega} _{+} }\le0$ (Please Refer to Appendix \ref{appendix C} for detailed expressions of $\varepsilon _{\bar{t}_{-}}$ and $\varepsilon _{\bar{\omega}_{+}}$). Also, we compute the necessary and sufficient condition, $\mathrm{Tr}[\rho[L_{\bar{t}_{-}},L_{\bar{\omega}_{+}}]]=0$.
	The joint estimation for the time difference and frequency sum, which is shown as
	\begin{eqnarray}
		\delta \bar{t}_{-}\delta \bar{\omega}_{+}\ge1.
	\end{eqnarray}
	\par
	The bound on the estimated time difference and frequency sum of the two single-photon states are much higher than those of the biphoton entangled state, and this advantage will be lost when $\kappa \le 0$. \par

	Moreover, we can achieve precise measurements of the size and velocity of the moving target by employing quantum illumination-based strategy. This approach also requires two pairs of entangled photons. Detailed discussion methods are provided in the appendix for reference. In this scheme, the joint estimation for $\bar{t}_{-}$ and $\bar\omega_{+}$ satisfies
	\begin{eqnarray}
		\delta \bar{t}_{-}\delta \bar{\omega}_{+}\ge  2\sqrt{1-\kappa^2}.\label{11}
	\end{eqnarray}
	It is imply that the uncertainty of simultaneously estimating $\bar{t}_{-}$ and $\bar\omega_{+}$ can be reduced by factor $2\sqrt{1-\kappa^2}$.\par

	As depicted in Figure \ref{figure 2}(b), we can compare the estimation accuracy of the size and velocity of moving objects for various types of detection sources and strategies. To ensure fairness, we maintain consistency by employing an equal number of photons for different probing strategies. The lower bound of the uncertainty of simultaneously estimating $\bar{t}_{-}$ and $\bar\omega_{+}$, denoted as $\mathrm{Min}[\delta \bar{t}_{-}\delta \bar{\omega}_{+}]$, are the functions of the parameter $\kappa$ for different strategies. In our strategy, we achieve $\mathrm{Min}[\delta \bar{t}_{-}\delta \bar{\omega}_{+}]=\frac{\sqrt{1-\kappa}}{\sqrt{1+\kappa}}$ (blue), while the quantum illumination scheme yields $\mathrm{Min}[\delta \bar{t}_{-}\delta \bar{\omega}_{+}]=2\sqrt{1-\kappa^2}$ (green). In the single-photon strategy, $\mathrm{Min}[\delta \bar{t}_{-}\delta \bar{\omega}_{+}]$ remains constant at $1$ throughout.

	It is clearly shown that, when utilizing a positively time-correlated entanglement source for estimation, $\kappa\in(0,1)$, our scheme proves to be optimal. However, for a negatively time-correlated entanglement source, the situation varies depending on the quality of the entanglement. Our strategy and the quantum illumination strategy fall short of the performance achieved by the single-photon strategy, when the entanglement is relatively weak, $\kappa\in(-\frac{\sqrt{3}}{2},0)$. Conversely, when the entanglement source approaches ideal negatively correlated in time, $\kappa\in(-1,-\frac{\sqrt{3}}{2})$, the the quantum illumination strategy emerges as the optimal solution similarly.
	
	\section{conclusion} \label{IV}
	
	We presented a quantum strategy for simultaneous estimation of two distinct physical quantities in various application scenarios, utilizing different quantum entanglement resources. Specifically, we explore the utilization of photons with positive or negative time correlation.
	
	The first aspect of our work focuses on proposing a method for detecting the central position and relative velocity of a multibody system through the measurement of time sum and frequency difference. Additionally, we present a technique for precise measurement of the size and velocity of moving objects.
	
	To provide a comprehensive analysis, we investigate the accuracy limits of the estimation parameters mentioned above, comparing them with the quantum illumination strategy and the single-photon strategy discussed earlier. Notably, our results highlight that when appropriate entanglement sources are utilized, and the entanglement degree is relatively high, the proposed quantum strategy exhibits a superior quantum advantage over the other two strategies.
	
	Overall, this study sheds light on the potential of leveraging quantum entanglement in simultaneous estimation tasks, demonstrating its superiority under certain conditions. These findings contribute to advancing our understanding of quantum strategies for accurate measurements in diverse scenarios.

	\section{Acknowledgment}
	
	We appreciate Lorenzo Maccone for his valuable comments and discussions. C.R. was supported by the National Natural Science Foundation of China (Grant No. 12075245, 12247105), the Natural Science Foundation of Hunan Province (2021JJ10033), Xiaoxiang Scholars Programme of Hunan Normal university.

	\bibliographystyle{apsrev4-1}
	\bibliography{ref}
	\clearpage
	\newpage
	\section{appendix}
	\subsection{Quantum Fisher Information}\label{appendix A}
	
	We review the basic concepts of quantum Fisher information. In many cases, the estimation of certain physical parameters relies on the association with another covariate. Such as, the position of a target can be estimated based on the travel time of single photons. In the context of unbiased estimators, the Quantum Cramér-Rao (QCR) bound determines the precision of parameter estimation,
	\setcounter{equation}{0}
	
	\renewcommand{\theequation}{A.\arabic{equation}}
	\begin{eqnarray}
		\delta \lambda ^{2}  & \ge  & \frac{1}{NH},
	\end{eqnarray}
	where $\lambda$ represents the estimated parameter, $H$ denotes the QFI matrix, and $N$ corresponds to the number of measurement. The QFI matrix is given by
	\begin{eqnarray}
		H=\frac{1}{2}Tr\left \{  \rho({ L_{\lambda _{i} }L_{\lambda _{j} }+L_{\lambda _{j} }L_{\lambda _{i} }})\right \},
	\end{eqnarray}
	where $\rho  =  \sum \rho_{ij} \lvert  e_{i}\rangle\langle e_{j} \rvert $  is the density matrix of the state expanded in an orthogonal basis, $L_{\lambda_{i}}$ and $L_{\lambda_{j}}$ are the symmetric logarithmic derivative(SLD)\cite{paris2009quantum}.
	
	The SLD matrix can be given as
	\begin{eqnarray}
		L_{\lambda_{ij}}=2\frac{  \langle e_{i} \lvert \partial _{\lambda } \rho  \lvert  e_{j}     \rangle  }{\rho_{ii}+\rho_{jj}}, \label{40}
	\end{eqnarray}
	where $\rho_{ii}$ and $\rho_{jj}$ are the element of the desity matrix, and $L_{\lambda_{ij}}$ is the element of SLD matrix.
	If the joint estimation of multiparameter is considered, a necessary and sufficient condition for the joint and optimal estimation is that the SLD operators commute, i.e.,
	\begin{eqnarray}
		Tr [\rho[L_{\lambda_{i}},L_{\lambda_{j}}]]=0.
	\end{eqnarray}
	
	If the estimated parameter is not the expected parameter，the correspondence between covariate and parameter can be given as,
	\begin{eqnarray}
		L_{\bar{\gamma}}=2\frac{  \langle e_{i} \lvert 	\frac{\partial {\rho   } }{\partial \bar{\lambda}}  	\frac{\partial {\bar{\lambda}    } }{\partial \bar{\gamma}  }  \lvert  e_{j}     \rangle  }{\rho_{ii}+\rho_{jj}},	\label{21}
	\end{eqnarray}
	where $\bar{\gamma}$ and $\bar{\lambda}$ are the actual estimated parameter and the covariate respectively. In this discussion, we use time and frequency to estimate distance and velocity respectively. The partial derivatives are
	\begin{flalign}
		\begin{aligned}
			& \ \frac{\partial \bar{t} }{\partial r}  =  \frac{2}{c(1-\Gamma) }
			& \ &
			\frac{\partial \bar{t } }{\partial \Gamma }  =  \frac{2r}{(1-\Gamma )^{2} }
			&& \\
			& \
			\frac{\partial \bar{\omega  } }{\partial r }  =  0
			& \ & \frac{\partial \bar{\omega  } }{\partial \Gamma  } =- \frac{2\bar{\omega }_{0}  }{(1-\Gamma )^{2} }
			&& \\
			& \ \frac{\partial {\sigma   } }{\partial r  }  = 0
			& \ & 	\frac{\partial {\sigma   } }{\partial \Gamma  } =- \frac{2{\sigma  }_{0}  }{(1-\Gamma )^{2} }
		\end{aligned},
	\end{flalign}
	with $\Gamma=\frac{v}{c}$, which has been given in \cite{PRXQuantum.2.030303}.
	The SLD matrix of velocity and position and the QFI matrix can be further written by correspondence of the above.
	
	\subsection{The reflected Photon state in time domain} \label{appendix B}
	
	Assuming that the distance between the moving target and the radar at time $t_0=0$ is $r$, the photon emitted at time $t$ reflects by the target, the photon will return at the time,
	\setcounter{equation}{0}
	
	\renewcommand{\theequation}{B.\arabic{equation}}
	\begin{eqnarray}
		\tau  =t+\frac{2r+2vt}{c-v}=t+\frac{2r}{c-v}+\frac{2v t}{c-v},\label{28}
	\end{eqnarray}
	where $v$ is the velocity of the moving target. Due to the Doppler effect, the frequency will change as
	\begin{eqnarray}
		\bar{\omega}   =  \frac{c-v}{c+v} \bar{\omega}_{0},\label{29}
	\end{eqnarray}
	where $\bar{\omega}$ is the carrier frequency of returned photons.\par
	
	If two moving targets with the radial velocities are $v_{1}$ and $v_{2}$ respectively, we assume that the two photons encountered the two targets and returned to the radar. The two photons emitted at $t_{1}$ and $t_{2}$, will return to the radar at $\tau_{i}=t_{i}+\frac{2r_{i}}{c-v_{i}}+\frac{2v_{i} t_{i}}{c-v_{i}}$, with i=1,2. Without loss of generality, the returned photons are described by the state
	\begin{eqnarray}
		\left | \psi^{'}  \right \rangle =\int \int \phi (\tau_{1},\tau_{2}  )\left | \tau_{1}   \right \rangle \left | \tau_{2}   \right \rangle d\tau_{1}d\tau_{2}.\label{30}
	\end{eqnarray}
	Combining Eqs.(\ref{28}), (\ref{29}) and (\ref{30}), we obtain Eq. (\ref{4}) shown in the main text.\par
	For two independent single-photon states, it can be calculated similarly. If the
	two photons are back-scattered by the targets, they will return
	with a time delay given by Eq.(\ref{28}). However, since the scattering source is incoherent, the returned two-photon state is a mixed state as defined in Eq.(\ref{8}).    \par

	\subsection{Estimates of the centeal position and relative velocity of multibody system} \label{appendix C}
	
	\subsubsection{Our strategy}
	
	The returned biphoton entangled state is Eq.(\ref{4}).
	To obtain the QFI matrix, we firstly derive the orthogonal basis
	\setcounter{equation}{0}
	\renewcommand{\theequation}{C.\arabic{equation}}
	\begin{eqnarray}
		\left | e_{n}   \right \rangle & = & \int e_{n}(t_{+} ,t_{-} )\left | t_{+}   \right \rangle \left |t_{-}  \right \rangle dt_{+} dt_{-},
	\end{eqnarray}
	where $n=1,2,3$,  and the temporal distribution can be expressed as
	\begin{numcases}{}
		\begin{aligned}        \nonumber
			e_{1}(t_{+} ,t_{-} )=\phi
		\end{aligned}  \\
		\begin{aligned}          \nonumber
			&e_{2}(t_{+} ,t_{-} )=2\sqrt{\frac{1}{\sigma _{1}^{2}+\sigma _{2}^{2}-2\kappa \sigma _{1}\sigma _{2} } }
			\\&\times [(t_{+}+t_{-}-\bar{t}_{+}-\bar{t}_{-}  )\sigma _{1}^{2}-2\kappa\sigma _{1}\sigma _{2}(t_{+}-\bar{t}_{+})
			\\&+\sigma _{2}^{2}(t_{+}-t_{-}-\bar{t}_{+}+\bar{t}_{-})]
			\phi  \\
		\end{aligned}\\
		\begin{aligned}
			&e_{3}(t_{+} ,t_{-} )=2\sqrt{\frac{(1-\kappa^{2})\sigma _{1}^{2}\sigma _{2}^{2}  }{\sigma _{1}^{2}+\sigma _{2}^{2}-2\kappa\sigma _{1}\sigma _{2}} } (t_{-}- \bar{t}_{-}  )\phi.\\
		\end{aligned}
	\end{numcases}
	
	Secondly, the SLDs for the estimation of the
	time sum $\bar{t}_{+}$ and frequency difference $\bar{\omega}_{-}$ can be derived,
	\begin{eqnarray}
		L_{_{\bar{t}_{+} } }=
		\begin{pmatrix}
			0 & \sqrt{\sigma _{1}^{2}+ \sigma _{2}^{2}-2\kappa\sigma _{1}\sigma _{2} }    & 0 \\
			\sqrt{\sigma _{1}^{2}+ \sigma _{2}^{2}-2\kappa\sigma _{1}\sigma _{2} }  & 0 & 0 \\
			0 &  0&  0\\
		\end{pmatrix} \nonumber
	\end{eqnarray}
	and
	\begin{eqnarray}	
		\begin{aligned}
			L_{_{\bar{\omega} _{-} } }&=\\
			&\begin{pmatrix}
				0 & -\frac{i}{2}\sqrt{\frac{\sigma _{1}^{2}-2\kappa\sigma _{1}\sigma _{2}+\sigma _{2}^{2}}{(1-\kappa^{2})\sigma _{1}^{2}\sigma _{2}^{2}}}    & 0 \\
				\frac{i}{2}\sqrt{\frac{\sigma _{1}^{2}-2\kappa\sigma _{1}\sigma _{2}+\sigma _{2}^{2}}{(1-\kappa^{2})\sigma _{1}^{2}\sigma _{2}^{2}}}& 0 & 0 \\
				0 &  0&  0\\
			\end{pmatrix}.
		\end{aligned}
	\end{eqnarray}
	Therefore, the QFI matrix is
	\begin{eqnarray}
		\begin{aligned}
			H (\bar{t}_{+},\bar{\omega}_{-})& =  \\
			&\begin{pmatrix}
				\sigma _{1}^{2}-2\kappa\sigma _{1}\sigma _{2}+\sigma _{2}^{2}   & 0 \\
				0  &  \frac{\sigma _{1}^{2}-2\kappa\sigma _{1}\sigma _{2}+\sigma _{2}^{2}}{4(1-\kappa^{2})\sigma _{1}^{2}\sigma _{2}^{2}}
			\end{pmatrix}.
		\end{aligned}	
	\end{eqnarray}
	Finally, we obtain the relationship
	\begin{eqnarray}
		\delta \bar{t}_{+}\delta \bar{\omega}_{-}\ge \sqrt{\frac{1}{H _{\bar{t}_{+}}H_ {\bar{\omega}_{-}}}}=\frac{2\sqrt{1-\kappa^{2}}\sigma _{1}\sigma _{2}}{\sigma _{1}^{2}-2\kappa\sigma _{1}\sigma _{2}+\sigma _{2}^{2}}.
	\end{eqnarray}
	As $\sigma_{1} \approx \sigma_{2}$, the relationship is approximately $\delta \bar{t}_{+}\delta \bar{\omega}_{-}\ge \sqrt{\frac{1+\kappa}{1-\kappa }}$.

	\subsubsection{The single photon strategy}
	
	For two single-photon states, the returned photons are incoherent and can be described by Eq.(\ref{8}).
	The orthogonal basis $\left | e_{n}  \right \rangle$ can be derived as
	\begin{numcases}{}
		\begin{aligned}      \nonumber
			\left | e_{1}  \right \rangle & = & \sqrt{\frac{1}{c_{1}}} (\left | \psi_{1}  \right \rangle +e^{i\bar{t}_{-}\bar{\omega}_{-} }\left | \psi_{2}  \right \rangle )
		\end{aligned}  \\
		\begin{aligned}  \nonumber
			\left | e_{2}  \right \rangle & = & \sqrt{\frac{1}{c_{2}}} ( \left | \psi_{1} \right \rangle -e^{i\bar{t}_{-}\bar{\omega}_{-} }\left | \psi_{2} \right \rangle )  \\
		\end{aligned}\\
		\begin{aligned}     \nonumber
			\left | e_{3}  \right \rangle =\sqrt{\frac{1}{c_{3}}}(\left | \partial_{\bar{t}_{-}} e_{1}  \right \rangle-\left \langle e_{1}  | \partial_{\bar{t}_{-}} e_{1}  \right \rangle\left | e_{1}  \right \rangle )   \\
		\end{aligned}\\
		\begin{aligned}
			\left | e_{4}  \right \rangle    =
			\sqrt{\frac{1}{c_{4}}}(\left | \partial_{\bar{t}_{-}} e_{2}  \right \rangle-\left \langle e_{2}  | \partial_{\bar{t}_{-}} e_{2}  \right \rangle\left | e_{2}  \right \rangle )
		\end{aligned},	\label{44}
	\end{numcases}
	where $c_{1}$, $c_{2}$, $c_{3}$ and $c_{4}$ are normalization factor. The mixed state Eq.(\ref{8}) can be expressed in orthogonal basis as
	\begin{eqnarray}
		\rho_{1}  = C_{1}  (\lvert  e_{1}  \rangle  \langle e_{1} \rvert  +C_{2}\lvert  e_{2}  \rangle  \langle e_{2} \rvert),\label{45}
	\end{eqnarray}
	with $C_{1}=(1+\frac{\sqrt{2\sigma_{1}\sigma_{2}}e^{\frac{-\bar{\omega}_{-}^{2}-4\bar{t}_{-}^{2}\sigma_{1}^{2}\sigma_{2}^2}{4(\sigma_{1}^{2}+\sigma_{2}^{2})}}}{\sqrt{\sigma_{1}^2+\sigma_{2}^2}}) $ and $C_{2}=(1-\frac{\sqrt{2\sigma_{1}\sigma_{2}}e^{\frac{-\bar{\omega}_{-}^{2}-4\bar{t}_{-}^{2}\sigma_{1}^{2}\sigma_{2}^2}{4(\sigma_{1}^{2}+\sigma_{2}^{2})}}}{\sqrt{\sigma_{1}^2+\sigma_{2}^2}}) $. \par
	
	According to Eq.(\ref{40}), we can obtain that
	\begin{eqnarray}
		L_{\bar{t}_{+}}=
		\begin{pmatrix}
			0 &2 a_{12}&0  &  2a_{14}  \\
			2 a_{21}  & 0 &  2a_{23} & 0  \\
			0 & 2a_{32} & 0 & 0 \\
			2a_{41} & 0 & 0 &  0
		\end{pmatrix},
	\end{eqnarray}
	with $a_{21}^{*}=a_{12}=C_{1}\left \langle \partial_{ \bar{t}_{+}}e_{1} | e_{2}  \right \rangle+C_{2}\left \langle \partial_{ \bar{t}_{+}}e_{2} | e_{1}  \right \rangle$,  $a_{41}^{*}=a_{14}=\left \langle \partial_{ \bar{t}_{+}}e_{1} | e_{4}  \right \rangle$, and   $a_{32}^{*}=a_{23}=\left \langle \partial_{ \bar{t}_{+}}e_{2}  | e_{3}  \right \rangle$.\par
	Similarly,
	\begin{eqnarray}
		L_{\bar{\omega}_{-}}=
		\begin{pmatrix}
			a_{11} & 0 & 2a_{13} &  0 \\
			0  & a_{22}  & 0 & 2a_{24}  \\
			2a_{31} & 0 & 0 & 0 \\
			0 & 2a_{42} & 0 &  0
		\end{pmatrix},
	\end{eqnarray}
	with $a_{11}=\frac{\partial_{ \bar{\omega}_{-}}C_{1}}{C_{1}}$,  $a_{31}^{*}=a_{13}=\left \langle \partial_{ \bar{\omega}_{-}}e_{1} | e_{3}  \right \rangle$, $a_{22}=\frac{\partial_{ \bar{\omega}_{-}}C_{2}}{C_{2}}$, and   $a_{42}^{*}=a_{24}=\left \langle \partial_{ \bar{\omega}_{-}}e_{2}  | e_{4}  \right \rangle$.
	
	So, the QFI matrix is
	\begin{eqnarray}
		H  =  \begin{pmatrix}
			H_{\bar{t}_{+}^2} & 0\\
			0 &  H_{\bar{\omega}_{-}^2}
		\end{pmatrix},
	\end{eqnarray}
	where $H_{\bar{t}_{+}^2}=\sigma_{1}^2+\sigma_{2}^2+\varepsilon _{\bar{t} _{+} } $
	, $H_{\bar{\omega}_{-}^2}=\frac{\sigma_{1}^2+\sigma_{2}^2}{4\sigma_{1}^2\sigma_{2}^2}+\varepsilon _{\bar{\omega} _{-} }$, $\varepsilon _{\bar{t} _{+} } <0$ and $\varepsilon _{\bar{\omega} _{-} }<0$.
	We calculated the necessary and sufficient condition for joint optimal estimation, $Tr[\rho[L_{\bar{t}_{+}},L_{\bar{\omega}_{-}}]]=0$.
	As $\sigma_{1} \approx \sigma_{2}$, we can determine that $H_{\bar{t}_{+}^2}=2\sigma^2-2e^{-\frac{\bar{\omega}_{-}^{2}+4\bar{t}_{-}^{2}\sigma^{4}}{4\sigma^{2}}}\bar{t}_{-}^{2}\sigma^{4} $,
	$H_{\bar{\omega}_{-}^2}=\frac{1}{2\sigma^2}-\frac{\bar{t}_{-}^{2}}{2}(-1+e^{\frac{\bar{\omega}_{-}^{2}+4\bar{t}_{-}^{2}\sigma^{4}}{4\sigma^{2}}})^{-1}$. The relationship between $\bar{t}_{+}$ and $\bar{\omega}_{-}$ is
	\begin{eqnarray}
		\delta \bar{t}_{+}\delta \bar{\omega}_{-}\ge 1.
	\end{eqnarray}

	\subsubsection{The quantum illumination strategy}
	
	For the quantum illumination strategy, the returned state can be described as
	\begin{eqnarray}
		\rho_{2}  =  \frac{1}{2}(\lvert  \Psi_{1}  \rangle  \langle \Psi_{1}\rvert  +\lvert  \Psi_{2} \rangle  \langle \Psi_{2} \rvert),
	\end{eqnarray}
	where $\lvert  \Psi_{i}  \rangle$ is defined in Eq.(\ref{4}) of the main text. \par
	
	The orthonormal basis is \cite{PRXQuantum.2.030303}
	\begin{numcases}{}
		\begin{aligned}      \nonumber
			\left | e_{1}  \right \rangle & = & \sqrt{\frac{1}{c_{1}}} (\left | \psi_{1}  \right \rangle +e^{i\bar{t}_{-}\bar{\omega}_{-} }\left | \psi_{2}  \right \rangle )
		\end{aligned}  \\
		\begin{aligned}  \nonumber
			\left | e_{2}  \right \rangle & = & \sqrt{\frac{1}{c_{2}}} ( \left | \psi_{1} \right \rangle -e^{i\bar{t}_{-}\bar{\omega}_{-} }\left | \psi_{2} \right \rangle )  \\
		\end{aligned}\\
		\begin{aligned}     \nonumber
			\left | e_{3}  \right \rangle =\sqrt{\frac{1}{c_{3}}}(\left | \partial_{\bar{t}_{-}} e_{1}  \right \rangle-\left \langle e_{1}  | \partial_{\bar{t}_{-}} e_{1}  \right \rangle\left | e_{1}  \right \rangle )   \\
		\end{aligned}\\
		\begin{aligned}
			\left | e_{4}  \right \rangle    =
			\sqrt{\frac{1}{c_{4}}}(\left | \partial_{\bar{t}_{-}} e_{2}  \right \rangle-\left \langle e_{2}  | \partial_{\bar{t}_{-}} e_{2}  \right \rangle\left | e_{2}  \right \rangle )  \nonumber
		\end{aligned}\\
		\begin{aligned}
			\left | e_{5}  \right \rangle    =
			\sqrt{\frac{1}{c_{5}}}&(\left | \partial_{\bar{\omega}_{-}} e_{1}  \right \rangle-\\
			&\left \langle e_{1}  | \partial_{\bar{\omega}_{-}} e_{1}  \right \rangle\left | e_{1}  \right \rangle-\left \langle e_{3}  | \partial_{\bar{\omega}_{-}} e_{1}  \right \rangle\left | e_{3}  \right \rangle )  \nonumber
		\end{aligned}\\
		\begin{aligned}
			\left | e_{6}  \right \rangle    =
			\sqrt{\frac{1}{c_{6}}}&(\left | \partial_{\bar{\omega}_{-}} e_{2}  \right \rangle-\\
			&\left \langle e_{2}  | \partial_{\bar{\omega}_{-}} e_{2}  \right \rangle\left | e_{2}  \right \rangle-\left \langle e_{4}  | \partial_{\bar{\omega}_{-}} e_{2}  \right \rangle\left | e_{4}  \right \rangle )
		\end{aligned},
	\end{numcases}
	where $c_{1}$, $c_{2}$, $c_{3}$, $c_{4}$, $c_{5}$ and $c_{6}$ are normalization factor. So the state can be diagonalized as
	\begin{eqnarray}
		\rho_{1}  = C_{1}  (\lvert  e_{1}  \rangle  \langle e_{1} \rvert  +C_{2}\lvert  e_{2}  \rangle  \langle e_{2} \rvert),
	\end{eqnarray}
	with $C_{1}=(1+e^{\frac{-\bar{\omega}_{-}^{2}-4\bar{t}_{-}^{2}\sigma^{4}}{8(1-\kappa^{2})\sigma^{2}}}) $ and $C_{2}=(1-e^{\frac{-\bar{\omega}_{-}^{2}-4\bar{t}_{-}^{2}\sigma^{4}}{8(1-\kappa^{2})\sigma^{2}}}) $.
	The SLD matrixs for $\bar{t}_{+}$ and $\bar{\omega}_{-}$ is
	\begin{eqnarray}
		L_{\bar{t}_{+}}  =  2\begin{pmatrix}
			0& a_{12} &0  &a_{14}  &0  &0 \\
			a_{21}& 0 & a_{23} & 0 & 0 &0 \\
			0 & a_{32} & 0 & 0 &0  &0 \\
			a_{41}& 0 &0  &0  &0  &0 \\
			0& 0 & 0 &0  & 0 & 0\\
			0 & 0 & 0 &  0&  0&0
		\end{pmatrix}
	\end{eqnarray}
	with $a_{21}^{*}=a_{12}=C_{1}\left \langle \partial_{ \bar{t}_{+}}e_{1} | e_{2}  \right \rangle+C_{2}\left \langle \partial_{ \bar{t}_{+}}e_{2} | e_{1}  \right \rangle$,  $a_{41}^{*}=a_{14}=\left \langle \partial_{ \bar{t}_{+}}e_{1} | e_{4}  \right \rangle$, and $a_{32}^{*}=a_{23}=\left \langle \partial_{ \bar{t}_{+}}e_{2}  | e_{3}  \right \rangle$, and
	\begin{eqnarray}
		L_{\bar{\omega}_{-}}  =  \begin{pmatrix}
			a_{11}& 0 &2a_{13}  &0  &2a_{15}  &0 \\
			0& a_{22} & 0 & 2a_{24} & 0 &2a_{26} \\
			2a_{31} &  & 0 & 0 &0  &0 \\
			0&2 a_{42} &0  &0  &0  &0 \\
			2a_{51}& 0 & 0 &0  & 0 & 0\\
			0 & 2a_{62} & 0 &  0&  0&0
		\end{pmatrix}
	\end{eqnarray}
	with $a_{11}=\frac{\partial_{ \bar{\omega}_{-}}C_{1}}{C_{1}}$,  $a_{31}^{*}=a_{13}=\left \langle \partial_{ \bar{\omega}_{-}}e_{1} | e_{3}  \right \rangle$, $a_{22}=\frac{\partial_{ \bar{\omega}_{-}}C_{2}}{C_{2}}$, $a_{42}^{*}=a_{24}=\left \langle \partial_{ \bar{\omega}_{-}}e_{2}  | e_{4}  \right \rangle$, $a_{51}^{*}=a_{51}=\left \langle \partial_{ \bar{\omega}_{-}}e_{1}  | e_{5}  \right \rangle$, and $a_{62}^{*}=a_{26}=\left \langle \partial_{ \bar{\omega}_{-}}e_{2}  | e_{6}  \right \rangle$.
	
	The QFI matrix is
	\begin{eqnarray}
		H  =  \begin{pmatrix}
			H_{\bar{t}_{+}^2} & 0\\
			0 &  H_{\bar{\omega}_{-}^2}
		\end{pmatrix},
	\end{eqnarray}
	where $H_{\bar{t}_{+}^2}=2\sigma^{2}+\varepsilon _{\bar{t} _{+}^{'} } $, $H_{\bar{\omega}_{-}^2}=\frac{\sigma^2}{2\sigma^2}+\varepsilon _{\bar{\omega} _{-} ^{'} }$, $\varepsilon _{\bar{t} _{+}^{'} } <0$ and $\varepsilon _{\bar{\omega} _{-}^{'} }<0$. $H_{\bar{t}_{+}^2}=2\sigma^2-2e^{-\frac{\bar{\omega}_{-}^{2}+4\bar{t}_{-}^{2}\sigma^{4}}{4\sigma^{2}}}\bar{t}_{-}^{2}\sigma^{4} $, $H_{\bar{\omega}_{-}^2}=\frac{1}{2(1-\kappa^2)\sigma^2}-\frac{\bar{t}_{-}^{2}}{2}(-1+e^{\frac{\bar{\omega}_{-}^{2}+4\bar{t}_{-}^{2}\sigma^{4}}{4(1-\kappa^2)\sigma^{2}}})^{-1}$. Finally, we can obtain the relationship,
	\begin{eqnarray}
		\delta \bar{t}_{+}\delta \bar{\omega}_{-}\ge 2\sqrt{1-\kappa^2}.
	\end{eqnarray}

	\subsection{Measurement the size and velocity of moving target}\label{appendix D}
	
	\subsubsection{Our strategy}
	
	The biphoton state received by the radar is Eq.(\ref{4}). We estimate two parameters($\bar{t}_{-}$, $\bar{\omega}_{+}$) by QFI.
	The orthonormal basis can be given by
		\setcounter{equation}{0}
	\renewcommand{\theequation}{D.\arabic{equation}}
	\begin{eqnarray}
		\left | e_{n}   \right \rangle & = & \int e_{n}(t_{+},t_{-})\left | t_{+}   \right \rangle \left |t_{-}  \right \rangle dt_{+} dt_{-}  ,
	\end{eqnarray}
	where $n=1,2,3$, the spectral distribution functions are
	\begin{numcases}{}
		\begin{aligned} \nonumber
			e _{1}(t_{+},t_{-})=\phi
		\end{aligned}  \\
		e_{2}(t_{+},t_{-})=\sqrt{2(1+\kappa )} \sigma (t_{-} -\bar{t}_{-})\phi  \nonumber \\
		e_{3}(t_{+},t_{-})=\sqrt{2(1-\kappa )} \sigma (t_{+} -\bar{t}_{+})\phi   .	
	\end{numcases}
	The SLD matrix are
	\begin{eqnarray}
		L_{_{\bar{t}_{-} } } & = & \begin{pmatrix}
			0 & \sqrt{2(1+\kappa )}\sigma   & 0 \\
			\sqrt{2(1+\kappa )}\sigma& 0 & 0 \\
			0 &  0&  0\\
			
		\end{pmatrix}
	\end{eqnarray}
	and
	\begin{eqnarray}
		L_{_{\bar{\omega} _{+} } } & = & \begin{pmatrix}
			0 & \frac{i}{2\sqrt{1-\kappa}\sigma  }    & 0\\
			\frac{-i}{\sqrt{2(1-\kappa)}\sigma  }& 0 & 0\\
			0 &  0&  0
		\end{pmatrix}.
	\end{eqnarray}
	The QFI matrix is
	\begin{eqnarray}
		H(\bar{t} _{-},\bar{\omega} _{+}) & = & \begin{pmatrix}
			2(1+\kappa )\sigma ^2 &0 \\
			0& \frac{1}{2(1-\kappa )\sigma ^2}
		\end{pmatrix}.
	\end{eqnarray}
	Therefore,
	\begin{eqnarray}
		\delta \bar{t}_{-}\delta \bar{\omega}_{+}\ge \sqrt{\frac{1-\kappa}{1+\kappa}}.
	\end{eqnarray}
	
	\subsubsection{The single photon strategy}
	
	For two single-photon states, similarly, the returned state and the basis vectors used to derive the QFI refer to Eq.\ref{44} and Eq.\ref{45}. The SLD matrix for $\bar{t}_{-}$ and $\bar{\omega}_{+}$ are
	\begin{eqnarray}
		L_{\bar{t}_{-}}=
		\begin{pmatrix}
			a_{11} & 0 & 2a_{13} &  0 \\
			0  & a_{22}  & 0 & 2a_{24}  \\
			2a_{31} & 0 & 0 & 0 \\
			0 & 2a_{42} & 0 &  0
		\end{pmatrix}, \nonumber
	\end{eqnarray}
	with $a_{11}=\frac{\partial_{ \bar{t}_{-}}C_{1}}{C_{1}}$,  $a_{31}^{*}=a_{13}=\left \langle \partial_{ \bar{t}_{-}}e_{1} | e_{3}  \right \rangle$, $a_{22}=\frac{\partial_{ \bar{t}_{-}}C_{2}}{C_{2}}$,  $a_{42}^{*}=a_{24}=\left \langle \partial_{ \bar{t}_{-}}e_{2}  | e_{4}  \right \rangle$, and
	\begin{eqnarray}
		L_{\bar{\omega}_{+}}=
		\begin{pmatrix}
			0 &2 a_{12}&0  &  2a_{14}  \\
			2 a_{21}  & 0 &  2a_{23} & 0  \\
			0 & 2a_{32} & 0 & 0 \\
			2a_{41} & 0 & 0 &  0
		\end{pmatrix},
	\end{eqnarray}
	with $a_{21}^{*}=a_{12}=C_{1}\left \langle \partial_{ \bar{\omega}_{+}}e_{1} | e_{2}  \right \rangle+C_{2}\left \langle \partial_{ \bar{\omega}_{+}}e_{2} | e_{1}  \right \rangle$,  $a_{41}^{*}=a_{14}=\left \langle \partial_{ \bar{\omega}_{+}}e_{1} | e_{4}  \right \rangle$, and   $a_{32}^{*}=a_{23}=\left \langle \partial_{ \bar{\omega}_{+}}e_{2}  | e_{3}  \right \rangle$.

	Then we can obtain the the QFI matrix,
	\begin{eqnarray}
		H  =  \begin{pmatrix}
			H_{\bar{t}_{-}^2} & 0\\
			0 &  H_{\bar{\omega}_{+}^2}
		\end{pmatrix},
	\end{eqnarray}
	with $H_{\bar{t}_{-}^2}=2\sigma^{2}-\frac{\bar{\omega}_{-}^{2}}{2}(-1+e^{\frac{\bar{\omega}_{-}^{2}+4\bar{t}_{-}^{2}\sigma^{2}}{4\sigma^{2}}})^{-1}$ and $H_{\bar{\omega}_{+}^2}=\frac{1}{2\sigma^2}-e^{-\frac{\bar{\omega}_{-}^{2}+4\bar{t}_{-}^{2}\sigma^{2}}{4\sigma^{2}}}\frac{\bar{\omega}_{-}^{2}}{2\sigma^4}$. The condition $Tr[\rho[L_{\bar{t}_{-}},L_{\bar{\omega}_{+}}]]=0$.\par
	
	The relationship of $\bar{t}_{-}$ and $\omega_{+}$ is
	\begin{eqnarray}
		\delta \bar{t}_{-}\delta \bar{\omega}_{+}\ge 1.
	\end{eqnarray}

	\subsubsection{The quantum illumination strategy}
	
	For quantum illumination strategy, the returned state is
	\begin{eqnarray}
		\rho_{2}  =  \frac{1}{2}(\lvert  \Psi_{1}  \rangle  \langle \Psi_{1}\rvert  +\lvert  \Psi_{2} \rangle  \langle \Psi_{2} \rvert),
	\end{eqnarray}
	where $\lvert  \Psi_{i}  \rangle$ refer to the main text Eq.(\ref{4}). \par
	the orthogonal basis is\cite{PRXQuantum.2.030303}
	\begin{numcases}{}
		\begin{aligned}      \nonumber
			\left | e_{1}  \right \rangle & = & \sqrt{\frac{1}{c_{1}}} (\left | \psi_{1}  \right \rangle +e^{i\bar{t}_{-}\bar{\omega}_{-} }\left | \psi_{2}  \right \rangle )
		\end{aligned}  \\
		\begin{aligned}  \nonumber
			\left | e_{2}  \right \rangle & = & \sqrt{\frac{1}{c_{2}}} ( \left | \psi_{1} \right \rangle -e^{i\bar{t}_{-}\bar{\omega}_{-} }\left | \psi_{2} \right \rangle )  \\
		\end{aligned}\\
		\begin{aligned}     \nonumber
			\left | e_{3}  \right \rangle =\sqrt{\frac{1}{c_{3}}}(\left | \partial_{\bar{t}_{-}} e_{1}  \right \rangle-\left \langle e_{1}  | \partial_{\bar{t}_{-}} e_{1}  \right \rangle\left | e_{1}  \right \rangle )   \\
		\end{aligned}\\
		\begin{aligned}
			\left | e_{4}  \right \rangle    =
			\sqrt{\frac{1}{c_{4}}}(\left | \partial_{\bar{t}_{-}} e_{2}  \right \rangle-\left \langle e_{2}  | \partial_{\bar{t}_{-}} e_{2}  \right \rangle\left | e_{2}  \right \rangle )  \nonumber
		\end{aligned}\\
		\begin{aligned}
			\left | e_{5}  \right \rangle    =
			\sqrt{\frac{1}{c_{5}}}&(\left | \partial_{\bar{\omega}_{-}} e_{1}  \right \rangle-\\
			&\left \langle e_{1}  | \partial_{\bar{\omega}_{-}} e_{1}  \right \rangle\left | e_{1}  \right \rangle-\left \langle e_{3}  | \partial_{\bar{\omega}_{-}} e_{1}  \right \rangle\left | e_{3}  \right \rangle )  \nonumber
		\end{aligned}\\
		\begin{aligned}
			\left | e_{6}  \right \rangle    =
			\sqrt{\frac{1}{c_{6}}}&(\left | \partial_{\bar{\omega}_{-}} e_{2}  \right \rangle-\\
			&\left \langle e_{2}  | \partial_{\bar{\omega}_{-}} e_{2}  \right \rangle\left | e_{2}  \right \rangle-\left \langle e_{4}  | \partial_{\bar{\omega}_{-}} e_{2}  \right \rangle\left | e_{4}  \right \rangle )
		\end{aligned},
	\end{numcases}
	where $c_{1}$, $c_{2}$, $c_{3}$, $c_{4}$, $c_{5}$ and $c_{6}$ are normalization factor. The state can be diagonalized as
	\begin{eqnarray}
		\rho_{1}  = C_{1}  (\lvert  e_{1}  \rangle  \langle e_{1} \rvert  +C_{2}\lvert  e_{2}  \rangle  \langle e_{2} \rvert),
	\end{eqnarray}
	with $C_{1}=(1+e^{\frac{-\bar{\omega}_{-}^{2}-4\bar{t}_{-}^{2}\sigma^{4}}{8(1-\kappa^{2})\sigma^{2}}}) $ and $C_{2}=(1-e^{\frac{-\bar{\omega}_{-}^{2}-4\bar{t}_{-}^{2}\sigma^{4}}{8(1-\kappa^{2})\sigma^{2}}}) $.
	The SLD matrix are
	\begin{eqnarray}
		L_{\bar{t}_{-}}  =  \begin{pmatrix}
			a_{11}& 0 &2a_{13}  &0  &0  &0 \\
			0& a_{22} & 0 & 2a_{24} & 0 &0 \\
			2a_{31} &  & 0 & 0 &0  &0 \\
			0&2 a_{42} &0  &0  &0  &0 \\
			0& 0 & 0 &0  & 0 & 0\\
			0 & 0 & 0 &  0&  0&0
		\end{pmatrix}
	\end{eqnarray}
	with $a_{11}=\frac{\partial_{ \bar{t}_{-}}C_{1}}{C_{1}}$,  $a_{31}^{*}=a_{13}=\left \langle \partial_{ \bar{t}_{-}}e_{1} | e_{3}  \right \rangle$, $a_{22}=\frac{\partial_{ \bar{t}_{-}}C_{2}}{C_{2}}$, $a_{42}^{*}=a_{24}=\left \langle \partial_{ \bar{t}_{-}}e_{2}  | e_{4}  \right \rangle$, $a_{51}^{*}=a_{51}=\left \langle \partial_{ \bar{t}_{-}}e_{1}  | e_{5}  \right \rangle$, $a_{62}^{*}=a_{26}=\left \langle \partial_{ \bar{t}_{-}}e_{2}  | e_{6}  \right \rangle$. And
	\begin{eqnarray}
		L_{\bar{\omega}_{+}}  =  2\begin{pmatrix}
			0& a_{12} &0  &a_{14}  &0  &a_{16} \\
			a_{21}& 0 & a_{23} & 0 & a_{25} &0 \\
			0 & a_{32} & 0 & 0 &0  &0 \\
			a_{41}& 0 &0  &0  &0  &0 \\
			0& a_{52} & 0 &0  & 0 & 0\\
			a_{61} & 0 & 0 &  0&  0&0
		\end{pmatrix}
	\end{eqnarray}
	with $a_{21}^{*}=a_{12}=C_{1}\left \langle \partial_{ \bar{\omega}_{+}}e_{1} | e_{2}  \right \rangle+C_{2}\left \langle \partial_{ \bar{\omega}_{+}}e_{2} | e_{1}  \right \rangle$,  $a_{41}^{*}=a_{14}=\left \langle \partial_{ \bar{\omega}_{+}}e_{1} | e_{4}  \right \rangle$, $a_{61}^{*}=a_{16}=\left \langle \partial_{ \bar{\omega}_{+}}e_{1}  | e_{6}  \right \rangle$,  $a_{32}^{*}=a_{23}=\left \langle \partial_{ \bar{\omega}_{+}}e_{2}  | e_{3}  \right \rangle$, $a_{52}^{*}=a_{25}=\left \langle \partial_{ \bar{\omega}_{+}}e_{2}  | e_{5}  \right \rangle$.
	
	The QFI matrix is
	\begin{eqnarray}
		H  =  \begin{pmatrix}
			H_{\bar{t}_{-}^2} & 0\\
			0 &  H_{\bar{\omega}_{+}^2}
		\end{pmatrix},
	\end{eqnarray}
	where $H_{\bar{t}_{-}^2}=2\sigma^{2}+\varepsilon _{\bar{t} _{-}^{'} } $, $H_{\bar{\omega}_{+}^2}=\frac{\sigma^2}{2\sigma^2}+\varepsilon _{\bar{\omega} _{+} ^{'} }$, $\varepsilon _{\bar{t} _{-}^{'} } <0$ and $\varepsilon _{\bar{\omega} _{+}^{'} }<0$. $H_{\bar{t}_{-}^2}=2\sigma^2-2e^{-\frac{\bar{\omega}_{-}^{2}+4\bar{t}_{-}^{2}\sigma^{4}}{4\sigma^{2}}}\frac{\bar{\omega}_{-}^{2}}{2} $, $H_{\bar{\omega}_{+}^2}=\frac{1}{2(1-\kappa^2)\sigma^2}-e^{\frac{\bar{\omega}_{-}^{2}+4\bar{t}_{-}^{2}\sigma^{4}}{4(1-\kappa^2)\sigma^{2}}}\frac{\bar{\omega}_{-}^{2}}{2\sigma^4}$. We can obtain the relationship,
	\begin{eqnarray}
		\delta \bar{t}_{-}\delta \bar{\omega}_{+}\ge 2\sqrt{1-\kappa^2}.
	\end{eqnarray}

	\clearpage
	\newpage
	
	
\end{document}